\documentclass[conference]{IEEEtran}
\IEEEoverridecommandlockouts
\usepackage{cite}
\usepackage{amsmath,amssymb,amsfonts}
\usepackage{algorithmic}
\usepackage{graphicx}
\usepackage{textcomp}
\usepackage{algorithm}
\usepackage{algorithmic}
\usepackage{xcolor}

\usepackage[left=1.62cm,right=1.62cm,top=0.7in]{geometry}

\newlength\myindent
\ifCLASSOPTIONcompsoc\usepackage[caption=false, font=normalsize, labelfont=sf, textfont=sf]{subfig}
\else
\usepackage[caption=false, font=footnotesize]{subfig}

\def\BibTeX{{\rm B\kern-.05em{\sc i\kern-.025em b}\kern-.08em
    T\kern-.1667em\lower.7ex\hbox{E}\kern-.125emX}}
\begin{document}

\title{Towards Mobility Management with Multi-Objective Bayesian Optimization\\
{\footnotesize}
\thanks{This work was supported by the European Union’s Horizon 2020 research and innovation programme under Marie Skłodowska-Curie Grant agreement No. 812991 (PAINLESS). G. Geraci was supported by grants RTI2018-101040, PID2021-123999OB-I00, and by the ``Ram\'{o}n y Cajal" program.}
}

\author{\IEEEauthorblockN{{Eloise~de~Carvalho~Rodrigues$^{\star \dagger}$, Alvaro~Valcarce~Rial$^{\star}$, and Giovanni~Geraci$^{\dagger}$}} \vspace{0.2cm}
\IEEEauthorblockA{$^{\star}$\emph{Nokia Bell Labs, Nozay, France}\\
\IEEEauthorblockA{$^{\dagger}$\emph{Universitat Pompeu Fabra (UPF), Barcelona, Spain}
}}}

\maketitle

\begin{abstract}
One of the consequences of network densification is more frequent handovers (HO). HO failures have a direct impact on the quality of service and are undesirable, especially in scenarios with strict latency, reliability, and robustness constraints. In traditional networks, HO-related parameters are usually tuned by the network operator, and automated techniques are still based on past experience.  In this paper, we propose an approach for optimizing HO thresholds using Bayesian Optimization (BO). We formulate a multi-objective optimization problem for selecting the HO thresholds that minimize HOs too early and too late in indoor factory scenarios, and we use multi-objective BO (MOBO) for finding the optimal values. Our results show that MOBO reaches Pareto optimal solutions with few samples and ensures service continuation through safe exploration of new data points.

\end{abstract}

%\begin{IEEEkeywords}
%Mobility, Industry 4.0, Bayesian Optimization, Handover parameters management
%\end{IEEEkeywords}

\section{Introduction}

The ongoing discussions on the directions of 5G---and beyond---suggest that future services and applications, such as Internet of Things (IoT), massive machine-type communications (mMTC), and the Industry 4.0, will require ever-increasing network capacity, high reliability and low latency, and robust mobility  \cite{Tayyab2020, Mollel2021}. With traffic heterogeneity and the consequent network densification, the coverage of each base station (BS) diminishes, leading to more frequent handovers (HO), i.e., transferring a user from one BS to another while maintaining connection. Those frequent HOs have a direct impact on the quality of service (QoS) and service continuity since HO problems are one of the main causes of service interruption in mobile networks \cite{Masri2021, 3gppData}.

Conventionally, a moving user equipment (UE) continuously monitors its signal quality and reports to its serving BS. In a nutshell, a HO occurs when the reference signal received power (RSRP) of the serving cell drops below an acceptable level and the RSRP from a neighbor BS (so-called target BS) is better than the serving cell's by a threshold \cite{Yajnanarayana2020}. This threshold, as well as other HO-related parameters, is usually tuned by the network operator based on human experience and domain knowledge. Mobility robustness optimization (MRO) techniques were introduced as an attempt to automate the selection of network parameters based on mobility key performance indicators (KPIs), but the methodology still relies on past observations and does not account for inconsistencies in the transmission environment. Improper selection of such HO parameters can lead to radio link failures (RLF) and unnecessary HOs, and it is difficult for trial-and-error schemes to achieve the required near-to-zero HO failures \cite{3gppData}.

Data-driven methods have drawn considerable attention to improve HO in future networks by using data for proactive decision-making with low management cost \cite{Mollel2021, Tanveer2022, sonmez2020handover, Tayyab2020, Yassein2017}. Recent works used machine learning (ML) techniques such as supervised learning \cite{ Farooq2022, Khan2021, Masri2021}, reinforcement learning (RL)  \cite{huang2022self, palas2021multi, Chang2021, Yajnanarayana2020}, and neural networks \cite{Lin2016, Ma2021} to optimize HO parameters and make mobility decisions. Bayesian optimization (BO) was used for coverage and capacity optimization \cite{Dreifuerst2020} and radio resource management (RRM) \cite{Maggi2020}.  Although mobility is a widely researched subject, there is still a lack of literature on industrial applications. The authors in \cite{hassan2019framework, lu2020feasibility, li2017cloud, gaur2020vertical, santi2022location} present HO management techniques for industrial IoT, but to the best of our knowledge, no current works apply multi-objective BO (MOBO) to mobility management and HO parameters optimization, and aside from \cite{Masri2021}, there are no other works considering indoor factory (InF) scenarios.

 In this paper, we vouch for MOBO to solve multi-objective RRM problems. BO is a promising technique since it can optimize costly functions with quick and efficient convergence, and exploration of new data points is done more safely when compared with non-probabilistic methods such as RL, thus avoiding drastic performance losses. Additionally, BO does not require a large amount of data to operate, addressing one of the main problems of ML in networks, i.e., the scarcity of data.

 In particular, we formulate a multi-objective problem to minimize both the number of HOs too late and too early in an InF scenario by choosing an optimal HO threshold. We use MOBO to solve the optimization problem and assess the performance of the method for different 3GPP InF channel model sub-scenarios. 
 Our results show that MOBO converges to a Pareto optimal solution within only 50 iterations for all considered scenarios while exploring a safe range of candidate samples. We also show the robustness of the learned models even with noisy data and give some insights for researchers interested in leveraging MOBO in mobility management problems.

\section{Problem Statement}
In this section, we detail the optimization problem formulation and provide an introduction to Bayesian Optimization and its multi-objective framework.

\subsection{Optimization Problem}

Wrong choices of HO parameters can lead to mobility problems and two of the most common issues that cause mobility interruptions are handovers happening \textit{too early} or \textit{too late}:

\begin{itemize}
    \item \textit{Handover too early:} happens when an RLF occurs shortly after a successful handover to the target cell, with the UE subsequently reconnecting to the source cell;

\item \textit{Handover too late:} happens when an RLF occurs at the serving cell before or during the handover procedure, with the UE subsequently reconnecting to the target cell (different from the source cell).
\end{itemize}

The above failures are directly linked to the choice of handover power thresholds, i.e., how better the target BS' RSRP needs to be for a UE to make a handover decision. Handovers too early are usually related to low HO thresholds, while handovers too late are generally related to higher HO thresholds. In practice, the HO procedure is a complex problem involving several parameters, including time-to-trigger, A1-A6 events, and hysteresis. In this work, for the sake of tractability and to better evaluate the performance of MOBO for mobility management, we focus on a single HO threshold $x$. We thus assume a HO to be triggered when a neighboring BS's RSRP is $x$ dB higher than the serving BS's RSRP.

The objective of the problem is to find a HO threshold $x$ that minimizes the total sum of HOs too early and HOs too late over a number of iterations in the network, as described in (\ref{eq1}).

\begin{equation}
\begin{aligned}
   \min\limits_{x} &  (f_1(x), f_2(x)) \\
     \text{where: } & f_1(x) = \sum \text{HO}_\text{too early}, \\
                   & f_2(x) = \sum \text{HO}_\text{too late}, \\
                   & x = \text{HO threshold [dB]}, \\
                   & x \in \{-5, -4, -3, ..., 10\} 
\end{aligned}
\label{eq1}
\end{equation}

We note that it is not possible to define a unified equation that relates the choice of networks parameters such as HO thresholds with the occurrence of HO failures, since these depend on complex network-environment interactions. Therefore, one way of tackling the above problem is through black-box optimization methods such as Bayesian Optimization.

\subsection{Bayesian Optimization in a Nutshell}

\begin{algorithm}[t]
\label{algobo}
\caption{Bayesian Optimization loop}
\begin{algorithmic}[1]
\STATE get initial samples $D_1$
\STATE initialize the model over the initial data
\STATE t=0
\WHILE{$t \leq T$}

    \STATE compute the acquisition function $\alpha(x)$---what are the points to be evaluated next? 
    \STATE compute $x_t = \arg \max_{x} \alpha (x|D_{1:t-1})$ 
    \STATE generate new samples by evaluating candidate points
    \STATE augment data set $ D_{1:t} = \{D_{1:t-1}, (x_t, f(x_t)) \}$ and update the model, $t=t+1$

\ENDWHILE{}
    
\STATE make a recommendation

\end{algorithmic}
\end{algorithm}

Bayesian Optimization \cite{shahriari2015taking} is a tool for maximizing expensive-to-evaluate black-box functions. Since we do not have access to the objective function $f$, the strategy is to treat $f$ as a random function and evaluate $f$ over a test sample set $D_1$ of $n$ initial observations, $D_1=\{(x_i, f(x_i)), i=1, 2, ..., n\}$, where $x_i$ are the input points and $f(x_i)$ are the output points. This prior captures the function's behavior and allows us to build a \textit{surrogate model} over the initial data samples. 

The surrogate model provides an \textit{estimation} of the actual $f$ in the form of a \textit{posterior distribution}, $p(f|D)$, where $D$ is a dataset with the complete history of samples and is used for evaluating new data samples and quantifying the uncertainty of the predictions. A commonly used surrogate model is a Gaussian Process (GP), defined in \cite{rasmussen2003gaussian} as  \textit{a collection of random variables, any finite number of which have a joint Gaussian distribution}, and is specified by a mean function $\mu(x)$ and a covariance kernel $k(x, x')$, with $x$ and $x'$ being any two points, such that \cite{rasmussen2003gaussian}:
\begin{equation}
\begin{aligned}
 & \mu(x) = \mathbb{E}[f(x)], \\
 & k(x, x') = \mathbb{E}[(f (x) - \mu(x)) (f (x') - \mu(x'))], \enspace \text{and} \\
 & f(x) \approx \text{GP}(\mu(x), k(x, x')).
 \end{aligned}
\end{equation}

Let us consider a new data sample $x_t$ and the past data samples $D_{1:t-1} = \{x_{1:t-1}, f(x_{1:t-1})\}$. The posterior distribution of $f({x_t})$ given the past data observations can be  computed following Bayes' rule\footnote{Bayes' rule can be described as the probability of an event given another prior related event, $p(a|b) = p(b|a)p(a)/p(b)$, where $p(a|b)$ is denoted the \textit{posterior}, $p(b|a)$ the \textit{likelihood}, $p(a)$ the \textit{prior}, and $p(b)$ the \textit{marginal likelihood}. For more details on deriving Bayes' rule for BO, see \cite{rasmussen2003gaussian}.}
\begin{equation}
    p(f(x_t)|x_t, D_{1:t-1}) = \mathcal{N}(\mu(x_t), \sigma^2(x_t))
\end{equation}
with $\sigma^2$ denoting the variance.

The covariance kernel indicates the uncertainty over a data point, i.e. points that are closer to each other are more likely to have similar function values than the ones farther apart. One can then \textit{explore} the high-uncertainty areas or \textit{exploit} low-variance candidates. The surrogate model is used for constructing an \textit{acquisition function} that  indicates the next point to be sampled. It is through the acquisition function that one can control the exploration/exploitation trade-off. Once a new candidate point is sampled, the dataset and the model are updated, and the optimization loop ends when it reaches a certain budget of $T$ function evaluations. The full BO loop is summarized in Algorithm 1. 

In multi-objective (MO) optimization, the goal is to find a set of Pareto optimal solutions, also called the \textit{Pareto front}. The Pareto front is a set of non-dominated solutions that are chosen to be optimal if, and only if, no objective can be improved without deteriorating another. Once the Pareto front is known, the decision-maker can use expert knowledge to determine the best solution based on the objectives trade-off.  MOBO offers a far more sample-efficient optimization when compared to other MO optimizers such as evolutionary algorithms \cite{Daulton2020}. 

In MOBO, the acquisition function is at the heart of the optimization loop, since it defines the points to be evaluated in the function being optimized.  One of the most recent and performance-efficient acquisition functions for MOBO is the $q$-Expected Hypervolume Improvement ($q$EHVI) \cite{Daulton2020}. While most of the acquisition functions for multi-objective rely on scalarization to turn the MO problem into a single objective or depend on heuristic approaches to expand sequential-evaluation algorithms to the parallel setting, qEHVI allows for parallel evaluation of $q$ new candidate points (instead of one at a time), and also provides the exact gradients of the Monte Carlo (MC) estimator so that gradient-based optimization methods can be employed. Another advantage of MOBO with $q$-EHVI is that, as it does not rely on scalarization, it makes it easier to determine the full set of Pareto optmal solutions. With the whole set of Pareto optimal solutions, one can switch from one solution to another without any intermediate performance loss.
\section{Simulation Setup}
\subsection{Indoor Factory Use-case}

\begin{figure}[t]
\centering
\includegraphics[width=0.9\columnwidth]{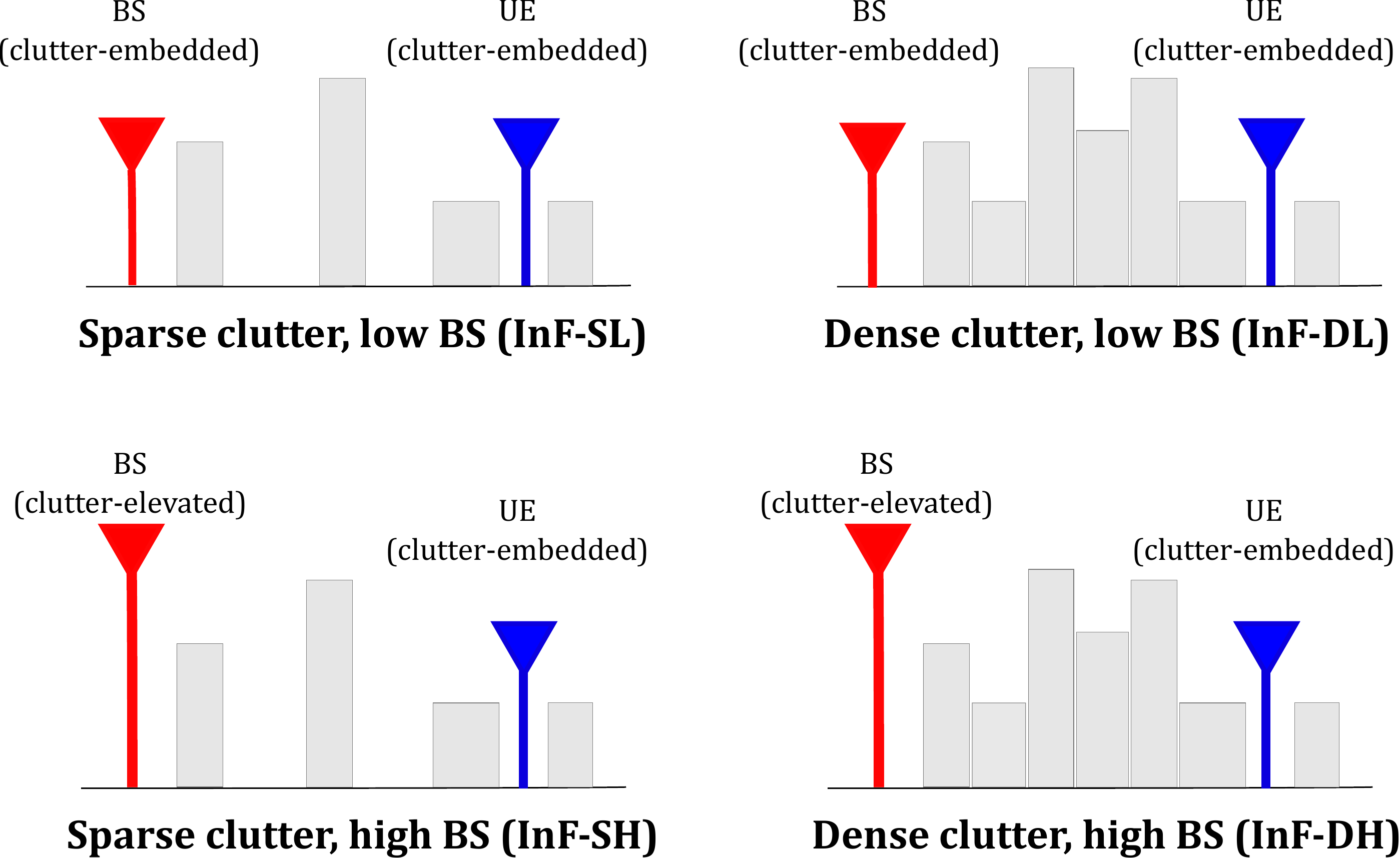}
\caption{3GPP Indoor Factory sub-scenarios.} 
\label{inf}
\end{figure}

With the rise of Industry 4.0, mobile stations have become a requisite for a smooth and flexible factory operation. In a future factory, for example, one can expect to have workers connected to the factory network, mobile robots being used for remote access and to provide connectivity in emergency cases, automated guided vehicles (AGV) for loading and unloading items, amongst dozens of other possible mobile stations. This mixed and flexible environment creates a challenging scenario for mobility optimization and may require sophisticated techniques. Each factory floor is unique and different from others. Hence, there are no one-size-fits-all mobility parameters, which is why site-specific mobility optimization is needed.

The 3GPP TR 38.901 \cite{3gppTR38901} proposes the Indoor Factory (InF) channel model that focuses on considering factory halls and the presence of ``clutter", e.g.: machinery of different sizes, assembly lines, and storage shelves, of various density and size. Unlike other 3GPP indoor models in which a non-line-of-sight (NLOS) condition is mainly caused by internal walls, people transiting, and office material, in InF, NLOS is defined as a function of the BS and UEs heights and clutter density. Depending on the height, the BS and UE can be considered as \textit{clutter embedded} or \textit{clutter elevated}, and depending on the density of clutter, the scenario can be considered as \textit{dense} or \textit{sparse}. Finally, there are four possible sub-scenarios to be considered as shown in Fig. \ref{inf}: Sparse clutter, low BS (InF-SL); Dense clutter, low BS (InF-DL); Sparse clutter, high BS (InF-SH); and Dense clutter, high BS (InF-DH), aside from the line-of-sight case with both UE and BS elevated.

\begin{table}[t]
    \centering
    \caption{InF sub-scenario specific parameters}
    \begin{tabular}{c|c|c|c|c|}
                    & InF-SL & InF-DL & InF-SH & InF-DH \\
         UE height $h_{UE}$ [m] &  1.5  &  1.5 &  1.5 & 1.5 \\
         BS height $h_{BS}$ [m]  &   1.5 &   1.5     & 8 & 8\\
         Clutter size  $d_\text{clutter}$ [m] &   10  & 2 & 10   & 2\\
         Clutter density $r$ [\%]  &  20 &  80 &  20  & 80\\
         Clutter height $h_c$ [m]  & 2 & 6  & 2 & 6\\
         Shadowing standard \\ deviation $\sigma$[dB] & 5.7 & 7.2  & 5.9 & 4.0\\
    \end{tabular}
    
    \label{tab:inf-tab}
\end{table}

The LOS probability determines the likelihood of a link to be in LOS or NLOS at a certain distance. The InF LOS probability is given as follows:
\begin{equation}
\Pr\{LOS\}=\exp{\left(-\frac{d_{2D}}{k}\right)},
\label{prlos}
\end{equation}
where
\begin{equation}
k=\left\{\begin{matrix}-\frac{d_{\text{clutter}}}{\ln{\left(1-r\right)}}&\mathrm{for\ InF-SL, InF-DL}\\
    -\frac{d_{\text{clutter}}}{\ln{\left(1-r\right)}} . \frac{h_{BS}-h_{UE}}{h_c-h_{UE}}&\mathrm{for\ InF-SH,InF-DH}\\\end{matrix}\right.
\end{equation}
and (\ref{prlos}) accounts for the blockage caused by machinery through the clutter size $d_{\text{cutter}}$, the clutter density $r$, and UE and BS antennas height $h_{UE}$ and $h_{BS}$, respectively.
Note that $k$ increases with the clutter density when the BS antenna is higher than the clutter, whereas the opposite occurs for a BS antenna lower than the clutter \cite{Jiang2021}. The scenario-specific parameters are summarized in Table \ref{tab:inf-tab}. Details on the LOS and NLOS pathloss models can be found in  \cite{3gppTR38901}.

\subsection{System Model}

\begin{figure}[t]
\centering
\includegraphics[width=0.65\columnwidth]{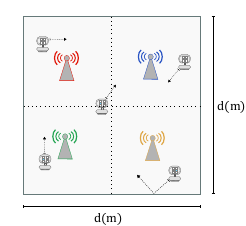}
\caption{Illustration of the Indoor Factory scenario considered.} 
\label{scenario}
\end{figure}

For assessing the gains obtained by the proposed MOBO methodology, we consider the $d \times d$ indoor factory scenario depicted in Fig. \ref{scenario}, with $d = 40~m$. This scenario contains ten mobile UEs randomly dropped in the area and four fixed-position BSs separated by a distance $d/2$. The height of BSs and UEs, as well as the clutter height, density, and size, all depend on the 3GPP-InF sub-scenario considered and are summarized in Table \ref{tab:inf-tab}.

Both UEs and BSs are equipped with one omnidirectional antenna and UEs associate with the BSs providing the highest received signal strength (RSS), which is recomputed each time the UE moves to the next position. As for the mobility model, we assume that UEs move on a straight line with a random departure angle, bouncing back when reaching the scenario's boundaries. We observe HO failures happening in a window of 250 s, with a time step of 0.5 s and a baseline UE velocity of 5 km/h. 

Pathloss and shadowing are modeled following the 3GPP InF standard. In this problem, mobility decisions are based on large-scale fading whereby the effect of small-scale fading is averaged out \cite{6384454}. The system operates at a carrier frequency of 3.5 GHz and the BS transmission power is set to 30 dBm, with a transmission bandwidth of 100 MHz \cite{3gppTR38901}. The main simulation parameters are listed in Table \ref{tab:param}.

\begin{table}[!t]
\caption{Simulation parameters \label{tab:param}}
\centering
\begin{tabular}{|c|c|}
\hline
\multicolumn{2}{|c|}{\text{\textbf{General parameters}}} \\
\hline
Monte Carlo samples &	300 \\
\hline
Number of UEs &	10 \\
\hline
Number of BSs	& 4 \\
\hline
Scenario size [m] &	40 \\
\hline
Channel model & 3GPP InF-DH \cite{3gppTR38901} \\
\hline
Noise spectral density [dB/MHz] & -174\\
\hline
Noise figure UEs [dB] & 9 \\
\hline
Bandwidth [MHz]	& 100 \\
\hline 
TX power BSs [dBm] & 30\\
\hline
$f_c$ [GHz] & 3.5\\
\hline
RFL power threshold [dBm] & -90\\
\hline
SINR RLF threshold [dB]	& -5\\
\hline
BS height [m] & 8\\
\hline
UE height [m] & 1.5\\
\hline
UE speed [km/h]	& 5\\
\hline
HO threshold [dB]&	[-5,…,10]\\
\hline
Mobility model &	Straight line with random angle\\
\hline\hline
\multicolumn{2}{|c|}{\text{\textbf{MOBO-specific parameters}}} \\
\hline
Reference point & [200, 200] \\
\hline
Number of full loops & 10 \\
\hline
Batch size & 4 points \\
\hline
Number of batches & 25 \\
\hline

\end{tabular}
\end{table}
 \section{Performance Evaluation}
In this section, we present and discuss our simulation results. As mentioned previously, the choice of HO parameters is usually made by the network operator based on expert knowledge. To simulate this approach as a benchmark, we consider an exhaustive search baseline in which HO thresholds from -5 dB to 10 dB are tested on each of the four InF sub-scenarios, and the number of HO failures over 250 seconds is observed. For statistical purposes, each HO threshold configuration is tested in 300 MC simulations. In each MC simulation, UEs are randomly re-dropped and moved with a fixed speed of 5 km/h. The MOBO algorithm is run using the BoTorch framework \cite{balandat2020botorch}, which is a library for Bayesian Optimization built atop PyTorch. We refer to a \textit{full optimization loop} as a complete run of MOBO, from initial sample generation to a final recommendation; the each full loop contains $b$ batches of $q$ candidate points to be evaluated in the surrogate model. In this problem, each objective is modeled as an independent GP with a Matérn 5/2 kernel, and the multi-objective problem is optimized using the $q$EHVI acquisition function.

 \begin{figure}[t]
     \centering
 \includegraphics[width=0.6\columnwidth]{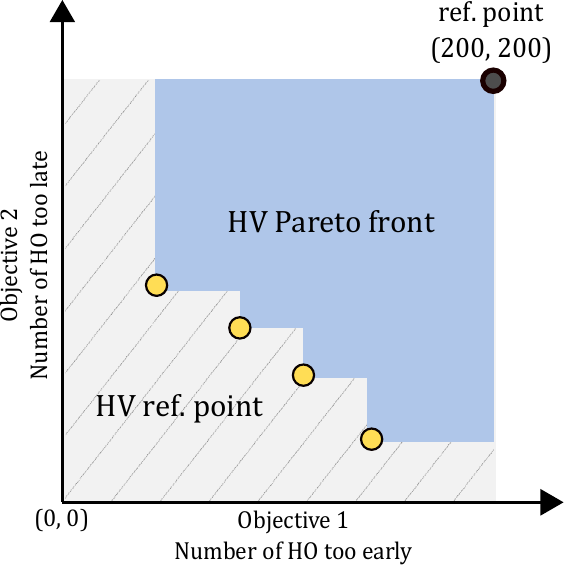}   \caption{Illustration of hypervolume considering the problem of minimizing both the number of HOs too early and too late and assuming a reference point of (200, 200) as being the maximum number of acceptable HO failures.}
 \label{fig:hv-pareto}
 \end{figure}

  \begin{figure*}[t]
    \centering
    \includegraphics[width=0.9\textwidth]{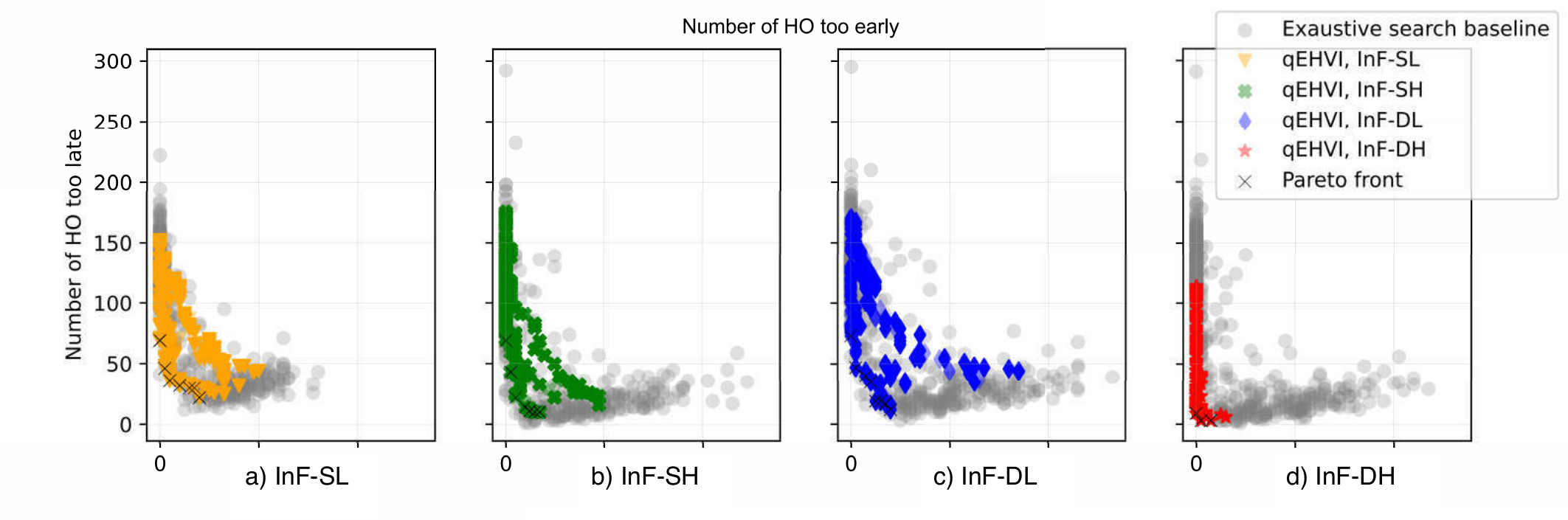}
    \caption{Complete search space with all samples obtained through exhaustive search vs the samples points by MOBO for the four channel models considered.}%a) InF-SL, b) InF-SH, c) InF-DL, and d) InF-DH.}
    \label{fig:bfbo}
\end{figure*}

  \begin{figure}[t!]
     \centering
 \includegraphics[width=0.95\columnwidth]{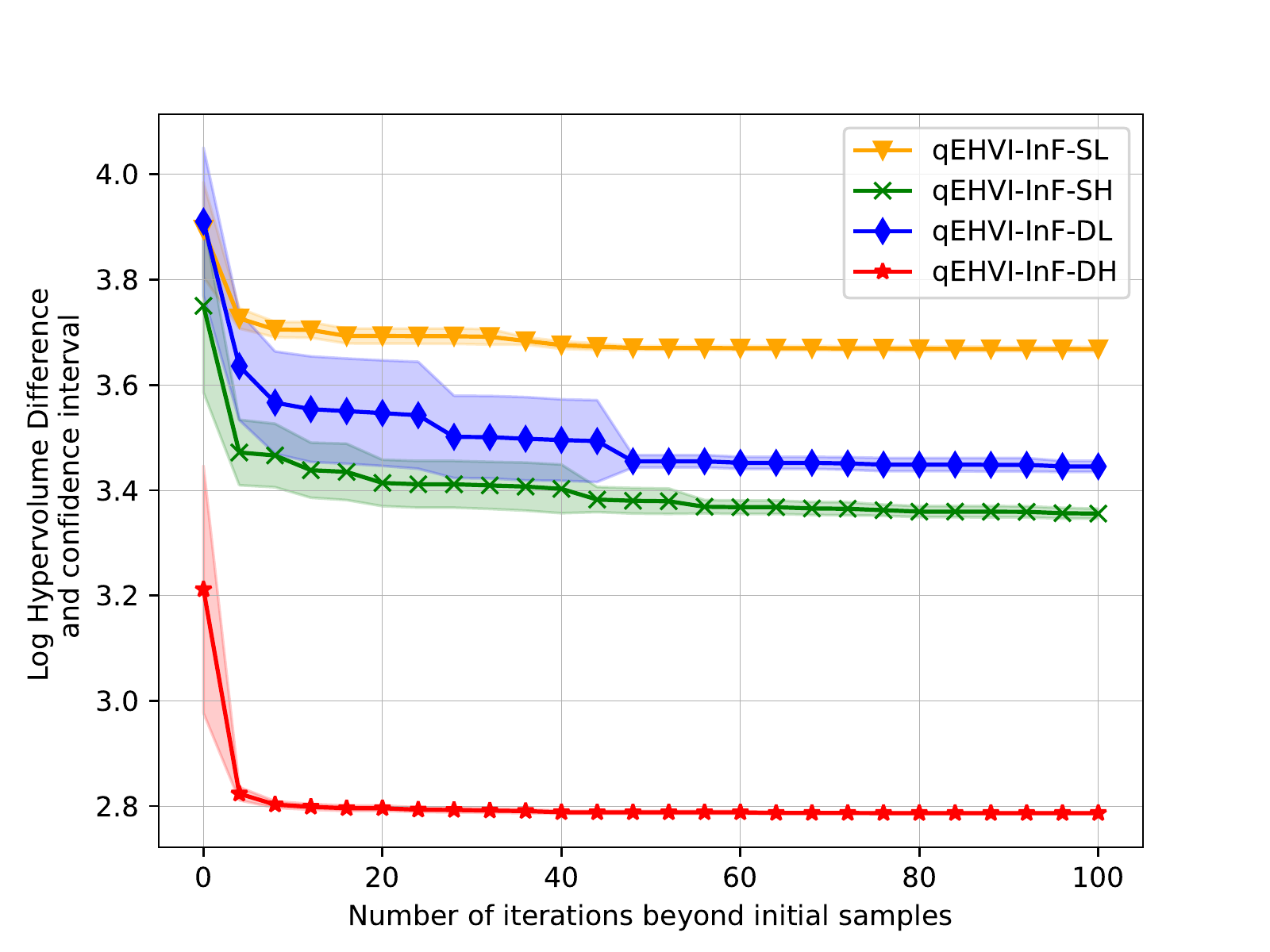}  
 \caption{Log HV difference of the solution found by MOBO-$q$EHVI for all of the InF sub-scenarios (markers) and confidence interval (shaded areas).}
 \label{fig:hv}
 \end{figure}

 A solid metric of the convergence and quality of a multi-objective problem's solution is the hypervolume (HV) \cite{riquelme2015performance}, i.e., the multi-dimensional volume of the region dominated by a set of solutions and bounded by a specified reference point. The log hypervolume difference is the log difference between the HV covered by the reference point and the HV of the approximated Pareto front obtained through MOBO as follows:
  \begin{equation}
     \log_{10} \text{HV}_\text{diff} = \log_{10} (\text{HV}_\text{ref. point} - \text{HV}_\text{MOBO}).
     \label{hveq}
 \end{equation}
In this paper, we consider as a reference point the maximum number of acceptable HO failures, namely 200 HOs too early/late, and define the latter as the bound in the MOBO algorithm. Fig. \ref{fig:hv-pareto} helps to visualize the HV calculation. Fig. \ref{fig:hv} shows the log hypervolume difference for the solutions found by the MOBO algorithm per iteration, averaged over 10 trials. The proposed algorithm reaches convergence in less than 50 iterations for all scenarios tested, determining the Pareto front.

  \begin{figure}[t!]
  \centering  
  \includegraphics[width=0.99\columnwidth]{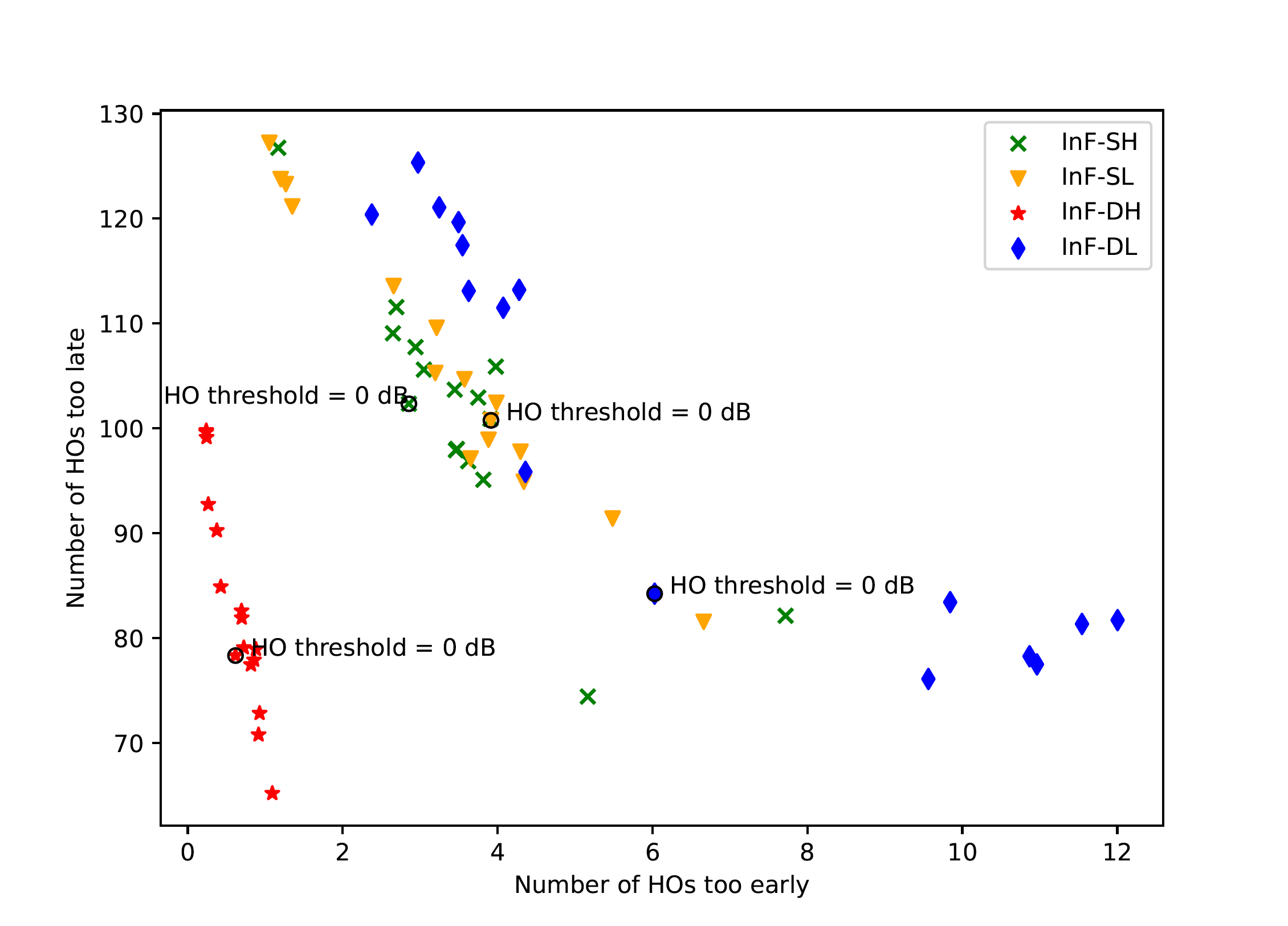}
  \caption{MOBO posterior model mean for all the InF sub-scenarios.}
    \label{fig:post3d}
 \end{figure}
 
 Fig. \ref{fig:bfbo} shows the full search space used by BO and the baseline exhaustive search approach. In this baseline, we observe all possible HO thresholds over 300 MC samples for each value, whereas MOBO runs 10 full optimization loops with $b$ = 25 batches of $q$ = 4 candidate points in each batch. The value of $q$ was chosen for the sake of complexity, as the optimization complexity increases with $q$. We note how MOBO samples points with lower variance, closer to the approximated Pareto front for all the InF sub-scenarios, and can find a sub-optimal solution using fewer samples than the baseline exhaustive search. The latter is able to eventually find solutions closer to zero-failures, but at the cost of requiring more samples and testing undesirable operating points that can dangerously jeopardize the system's performance. Although there is no guarantee that MOBO will never make a decision that can degrade the network performance, the search space in MOBO is safely restricted, i.e., worst-case limits can be set as parameters before the optimization with an error margin. In this problem, the limit was set to 200 HO failures, but was never reached by MOBO, showing its robustness and safe exploration.

Fig. \ref{fig:post3d} shows an average over 10 trials of the posterior models learned by the MOBO algorithm for each HO threshold tested. The posterior distribution is used for estimating the posterior value of the black-box function given a data sample, and this distribution is used to decide which point to sample next. Note that the models are only an \textit{approximation} of the true objective function and the MOBO algorithm focuses on sampling points closer to the approximate Pareto front, therefore the ``non-monotonic'' behavior in Fig. \ref{fig:post3d} can be caused by the optimization process's outliers. It is possible to reach a smoother curve by  increasing the number of trials, but we choose to limit the number of trials to showcase that MOBO only needs a few iterations for convergence and can operate even with noisy data samples. The models learned can also be used as a starting point for future optimization adjustments.

\section{Conclusions}
In this paper, we formulated a multi-objective optimization problem for determining the HO threshold that minimizes both the number of HOs too early and too late in various 3GPP indoor factory scenarios. We then used multi-objective Bayesian Optimization to solve the problem and evaluated the performance of the MOBO algorithm against a baseline exhaustive search. Our results showed that, although exhaustive search can reach a solution closer to zero, it does so at the cost of needing more sample points and exploring parameters that can drastically degrade the system's performance. Instead, MOBO reaches a Pareto optimal solution with fewer samples and guarantees a safer exploration, ensuring sustained QoS. 

We also showed that the model learned by MOBO has a high confidence interval even when using noisy data. 
While GP is usually the favorite choice of surrogate model, the problem at hand deals with discrete counters and its posterior model may be better estimated through a discrete surrogate model. %which would explain the noisy behavior of the data samples
A promising research direction is thus employing discrete models such as Random Forests or Beta-Bernoulli bandits.

In practical scenarios, implementing MOBO with $q$EHVI would entail the parallel evaluation of the multiple candidate points generated. This could be accomplished either (i) evaluating the points in parallel by creating subsets of users to test each of the candidate points, or (ii) testing the points sequentially using the entire network. The latter approach would still yield a performance advantage since $q$EHVI samples multiple candidate points at the same time, thus requiring fewer steps to maximize the acquisition function than a fully sequential optimization approach.

\bibliographystyle{IEEEtran}
\bibliography{bibliography.bib}

\begin{thebibliography}{10}
\providecommand{\url}[1]{#1}
\csname url@samestyle\endcsname
\providecommand{\newblock}{\relax}
\providecommand{\bibinfo}[2]{#2}
\providecommand{\BIBentrySTDinterwordspacing}{\spaceskip=0pt\relax}
\providecommand{\BIBentryALTinterwordstretchfactor}{4}
\providecommand{\BIBentryALTinterwordspacing}{\spaceskip=\fontdimen2\font plus
\BIBentryALTinterwordstretchfactor\fontdimen3\font minus
  \fontdimen4\font\relax}
\providecommand{\BIBforeignlanguage}[2]{{%
\expandafter\ifx\csname l@#1\endcsname\relax
\typeout{** WARNING: IEEEtran.bst: No hyphenation pattern has been}%
\typeout{** loaded for the language `#1'. Using the pattern for}%
\typeout{** the default language instead.}%
\else
\language=\csname l@#1\endcsname
\fi
#2}}
\providecommand{\BIBdecl}{\relax}
\BIBdecl

\bibitem{Tayyab2020}
M.~Tayyab, X.~Gelabert, and R.~J{\"a}ntti, ``{A survey on handover management:
  From LTE to NR},'' \emph{IEEE Access}, vol.~7, pp. 118\,907--118\,930, 2019.

\bibitem{Mollel2021}
M.~S. Mollel \emph{et~al.}, ``{A survey of machine learning applications to
  handover management in 5G and beyond},'' \emph{IEEE Access}, vol.~9, pp.
  45\,770--45\,802, 2021.

\bibitem{Masri2021}
A.~Masri \emph{et~al.}, ``Machine-learning-based predictive handover,'' in
  \emph{2021 IFIP/IEEE International Symposium on Integrated Network Management
  (IM)}.\hskip 1em plus 0.5em minus 0.4em\relax IEEE, 2021, pp. 648--652.

\bibitem{3gppData}
``{Evolved Universal Terrestrial Radio Access (E-UTRA) and NR; Study on
  enhancement for Data Collection for NR and EN-DC (Release 17)},'' \emph{3GPP
  TR 37.817}, January 2022.

\bibitem{Yajnanarayana2020}
V.~Yajnanarayana, H.~Ryden, and L.~Hevizi, ``{5G Handover using Reinforcement
  Learning},'' \emph{2020 IEEE 3rd 5G World Forum, 5GWF 2020 - Conference
  Proceedings}, pp. 349--354, 2020.

\bibitem{Tanveer2022}
J.~Tanveer, A.~Haider, R.~Ali, and A.~Kim, ``{An Overview of Reinforcement
  Learning Algorithms for Handover Management in 5G Ultra-Dense Small Cell
  Networks},'' \emph{Applied Sciences}, vol.~12, no.~1, 2022.

\bibitem{sonmez2020handover}
{\c{S}}.~S{\"o}nmez, I.~Shayea, S.~A. Khan, and A.~Alhammadi, ``Handover
  management for next-generation wireless networks: A brief overview,''
  \emph{IEEE Microwave Theory and Techniques in Wireless Communications},
  vol.~1, pp. 35--40, 2020.

\bibitem{Yassein2017}
M.~B. Yassein, S.~Aljawarneh, and W.~Al-Sarayrah, ``Mobility management of
  internet of things: Protocols, challenges and open issues,'' in \emph{Proc.
  International Conference on Engineering \& MIS}, 2017, pp. 1--8.

\bibitem{Farooq2022}
M.~U.~B. Farooq, M.~Manalastas, S.~M.~A. Zaidi, A.~Abu-Dayya, and A.~Imran,
  ``Machine learning aided holistic handover optimization for emerging
  networks,'' \emph{arXiv preprint arXiv:2202.02851}, 2022.

\bibitem{Khan2021}
S.~A. Khan \emph{et~al.}, ``An improved handover decision algorithm for {5G}
  heterogeneous networks,'' \emph{Proc. IEEE Malaysia International Conference
  on Communications}, pp. 25--30, 2021.

\bibitem{huang2022self}
W.~Huang, M.~Wu, Z.~Yang, K.~Sun, H.~Zhang, and A.~Nallanathan,
  ``{Self-Adapting Handover Parameters Optimization for SDN-Enabled UDN},''
  \emph{IEEE Transactions on Wireless Communications}, 2022.

\bibitem{palas2021multi}
M.~R. Palas \emph{et~al.}, ``{Multi-criteria handover mobility management in 5G
  cellular network},'' \emph{Computer Communications}, vol. 174, pp. 81--91,
  2021.

\bibitem{Chang2021}
S.~Chang \emph{et~al.}, ``{Individualized optimization of handover parameters
  in 5G (NR) indoor and outdoor co-frequency scenarios},'' \emph{Proc. IEEE
  ICCECE}, pp. 59--66, 2021.

\bibitem{Lin2016}
P.~C. Lin, L.~F. Casanova, and B.~K. Fatty, ``{Data-Driven Handover
  Optimization in Next Generation Mobile Communication Networks},''
  \emph{Mobile Information Systems}, vol. 2016, 2016.

\bibitem{Ma2021}
Y.~Ma, X.~Chen, and L.~Zhang, ``{Base Station handover Based on User Trajectory
  Prediction in 5G Networks},'' \emph{19th IEEE International Symposium on
  Parallel and Distributed Processing with Applications}, pp. 1476--1482, 2021.

\bibitem{Dreifuerst2020}
R.~M. Dreifuerst \emph{et~al.}, ``Optimizing coverage and capacity in cellular
  networks using machine learning,'' in \emph{Proc. IEEE ICASSP}, 2021, pp.
  8138--8142.

\bibitem{Maggi2020}
L.~Maggi, A.~Valcarce, and J.~Hoydis, ``Bayesian optimization for radio
  resource management: Open loop power control,'' \emph{IEEE Journal on
  Selected Areas in Communications}, vol.~39, no.~7, pp. 1858--1871, 2021.

\bibitem{hassan2019framework}
R.~Hassan, A.~H.~M. Aman, L.~A. Latiff \emph{et~al.}, ``{Framework for handover
  process using visible light communications in 5G},'' in \emph{Proc. Symposium
  on Future Telecommunication Technologies}, vol.~1, 2019, pp. 1--4.

\bibitem{lu2020feasibility}
Y.~Lu \emph{et~al.}, ``Feasibility of location-aware handover for autonomous
  vehicles in industrial multi-radio environments,'' \emph{Sensors}, vol.~20,
  no.~21, p. 6290, 2020.

\bibitem{li2017cloud}
D.~Li, X.~Li, and J.~Wan, ``A cloud-assisted handover optimization strategy for
  mobile nodes in industrial wireless networks,'' \emph{Computer Networks},
  vol. 128, pp. 133--141, 2017.

\bibitem{gaur2020vertical}
A.~S. Gaur, J.~Budakoti, C.-H. Lung \emph{et~al.}, ``Vertical handover decision
  for mobile iot edge gateway using multi-criteria and fuzzy logic
  techniques,'' \emph{Advances in Internet of Things}, vol.~10, no.~04, p.~57,
  2020.

\bibitem{santi2022location}
S.~Santi, F.~Lemic, and J.~Famaey, ``{Location-based Discovery and Network
  Handover Management for Heterogeneous IEEE 802.11 ah IoT Applications},''
  \emph{IEEE Trans. Network and Service Management}, 2022.

\bibitem{shahriari2015taking}
B.~Shahriari, K.~Swersky, Z.~Wang, R.~P. Adams, and N.~De~Freitas, ``Taking the
  human out of the loop: A review of bayesian optimization,'' \emph{Proceedings
  of the IEEE}, vol. 104, no.~1, pp. 148--175, 2015.

\bibitem{rasmussen2003gaussian}
C.~E. Rasmussen, ``Gaussian processes in machine learning,'' in \emph{Summer
  school on machine learning}.\hskip 1em plus 0.5em minus 0.4em\relax Springer,
  2003, pp. 63--71.

\bibitem{Daulton2020}
S.~Daulton, M.~Balandat, and E.~Bakshy, ``Differentiable expected hypervolume
  improvement for parallel multi-objective bayesian optimization,''
  \emph{Advances in Neural Information Processing Systems}, vol.~33, pp.
  9851--9864, 2020.

\bibitem{3gppTR38901}
\BIBentryALTinterwordspacing
(2019) {Study on channel model for frequencies from 0.5 to 100 GHz (Release
  16)}. [Online]. Available: \url{https://www.3gpp.org}
\BIBentrySTDinterwordspacing

\bibitem{Jiang2021}
T.~Jiang \emph{et~al.}, ``{3GPP standardized 5G channel model for IIoT
  scenarios: A survey},'' \emph{IEEE Internet of Things Journal}, vol.~8,
  no.~11, pp. 8799--8815, 2021.

\bibitem{6384454}
D.~Lopez-Perez, I.~Guvenc, and X.~Chu, ``Mobility management challenges in 3gpp
  heterogeneous networks,'' \emph{IEEE Communications Magazine}, vol.~50,
  no.~12, pp. 70--78, 2012.

\bibitem{balandat2020botorch}
M.~Balandat \emph{et~al.}, ``Botorch: a framework for efficient monte-carlo
  bayesian optimization,'' \emph{Advances in neural information processing
  systems}, vol.~33, pp. 21\,524--21\,538, 2020.

\bibitem{riquelme2015performance}
N.~Riquelme, C.~Von~L{\"u}cken, and B.~Baran, ``Performance metrics in
  multi-objective optimization,'' in \emph{2015 Latin American computing
  conference (CLEI)}.\hskip 1em plus 0.5em minus 0.4em\relax IEEE, 2015, pp.
  1--11.

\end{thebibliography}

\vspace{12pt}

\end{document}